\documentclass[%
 aip,
jap,%
 sd,%
 amsmath,amssymb,
 reprint,%
]{revtex4-1}

\draft % marks overfull lines with a black rule on the right

\usepackage{graphicx}% Include figure files
\usepackage{dcolumn}% Align table columns on decimal point
\usepackage{bm}% bold math
\usepackage{textcomp}
\usepackage[centerlast]{caption}
\usepackage{subcaption}
\usepackage{amssymb}
\usepackage{natbib}
\setcitestyle{numbers,open={[},close={]}}

\begin{document}

\preprint{AIP/123-QED}

\title[Sample title]{Highly confined Love waves modes by defect states\\ in a holey SiO$_{2}$/quartz phononic crystal\footnote{Accepted for publication in Journal of Applied Physics}}% Force line breaks with \\

\author{Yuxin Liu}
 \email{yuxin.liu@phd.ec-lille.fr}
 
\author{Abdelkrim Talbi}%

\author{Philippe Pernod}

\author{Olivier Bou Matar}%
\affiliation{Univ. Lille, CNRS, Centrale Lille, ISEN, Univ. Valenciennes, UMR 8520 IEMN, LIA LICS/LEMAC}%

%\author{XXX}

%\author{ }

\date{\today}% It is always \today, today,

\begin{abstract}
Highly confined Love modes are demonstrated in a phononic crystal based on a square array of etched holes in \textrm{SiO$_{2}$} deposited on the ST-cut quartz. An optimal choice of the geometrical parameters contributes to a wide stop-band for shear waves' modes. The introduction of a defect by removing lines of holes leads to the nearly flat modes within the band gap and consequently paves the way to implement advanced designs of electroacoustic filters and high-performance cavity resonators. The calculations are based on the finite element method in considering the elastic and piezoelectric properties of the materials. Interdigital transducers are employed to measure the transmission spectra. The geometrical parameters enabling the appearance of confined cavity modes within the band gap and the efficiency of the electric excitation were investigated. 
\end{abstract}


\maketitle

%\begin{quotation}
%The ``lead paragraph'' is encapsulated with the \LaTeX\ 
%\verb+quotation+ environment and is formatted as a single paragraph before %the first section heading. 
%(The \verb+quotation+ environment reverts to its usual meaning after the first sectioning command.) 
%Note that numbered references are allowed in the lead paragraph.
%
%The lead paragraph will only be found in an article being prepared for the %journal \textit{Chaos}.
%\end{quotation}

\section{INTRODUCTION}
Phononic crystals (PnCs), as an elastic analog of the photonic crystals, have received increasing attention in the last two decades. PnCs are widely investigated for their potential applications in various areas, including RF communications\cite{khelif_trapping_2003,khelif_experimental_2005,wu_waveguiding_2009,olsson_ultra_2009,pennec_two-dimensional_2010,liang_acoustic_2010,feng_micro-silicon_2017}, acoustic isolators\cite{meseguer_rayleigh-wave_1999,wu_layered_2008,olsson_iii_microfabricated_2009,ziaei-moayyed_silicon_2011,liu_design_2014,binci_planar_2016}, sensors\cite{talbi_zno/quartz_2006,lucklum_phononic_2009,ke_sub-wavelength_2011,zubtsov_2d_2012,salman_low-concentration_2015,xu_implementation_2018}, thermoelectric materials\cite{kim_thermal_2011,yu_reduction_2010,yang_extreme_2014,zen_engineering_2014} and meta-materials\cite{zhang_negative_2004,profunser_dynamic_2009,zhu_holey-structured_2011,narayana_heat_2012,zhao_beam_2014,li_acoustic_2015,tsai_manipulation_2016,page_focusing_2016,guo_acoustic_2017}. Composed of 1D, 2D, or 3D periodic arrays of inclusions embedded in a matrix, PnCs give rise to the complete or partial band gaps for both bulk acoustic waves (BAW)\cite{martinez-sala_sound_1995,liu_locally_2000,vasseur_experimental_2001,yang_ultrasound_2002,khelif_trapping_2003,pennec_two-dimensional_2010} and surface acoustic waves (SAW)\cite{dhar_high_2000,wu_frequency_2005,khelif_complete_2006,wu_layered_2008,mohammadi_evidence_2008,benchabane_observation_2011,achaoui_local_2013,liu_design_2014,liu_evidence_2014,hemon_hypersonic_2014}. The introduction of defects into PnCs is at the origin of multiple applications such as waveguide\cite{torres_ultrasonic_2001,khelif_experimental_2005,olsson_microfabricated_2008}, cavity\cite{khelif_trapping_2003,mohammadi_high-q_2009,jin_tunable_2016}, filter\cite{khelif_two-dimensional_2003,pennec_two-dimensional_2010} and multiplexer\cite{pennec_acoustic_2005}. Most research on the defect modes is based on the bulk waves\cite{khelif_trapping_2003,khelif_experimental_2005,pennec_two-dimensional_2010}, Rayleigh waves\cite{benchabane_guidance_2015} and Lamb waves\cite{miyashita_sonic_2005,mohammadi_high-q_2009}, while sensors, especially the bio-sensors, are based on the Love waves and antisymmetric Lamb waves, which are compatible with the liquid environment\cite{lucklum_phononic_2009,ke_sub-wavelength_2011} and leak less energy in the liquid. However, Lamb waves propagate on the extremely thin slabs, making them comparably fragile and therefore difficult to manipulate. Whereas Love waves, a shear horizontal (SH) polarized SAW, exist in the guiding layer deposited on a semi-infinite substrate, which guarantees both the confinement of the energy and the toughness of the device, in comparison with the Lamb waves devices. In recent years, the partial band-gap effect of PnCs on Love waves has been reported and a reflective grating was then proposed\cite{liu_evidence_2014,liu_design_2014}. Nevertheless, the exploitation of Love waves interacting with the defect states in PnCs remains to be investigated. 

In this paper, we demonstrate the acoustic band gap effect in a 2D PnC consisting of a square array of holes in a thin amorphous SiO$_{2}$ (silica) layer covering a ST-cut quartz substrate. Localized defect or cavity modes in the band gap which are introduced by removing lines of holes in the lattice are observed. 
The efficiency of cavity modes in the isolation of PnC 
is investigated as a function of the geometrical parameters of PnC and cavity.  
These effects are used to design micro-electromechanical resonators with highly confined cavity modes of Love waves. 
The band structures and transmission spectra are calculated with the finite element method (FEM, COMSOL Multiphysics$^{\circledR}$). The transmission spectra are compared with the dispersion curves and the resonant frequencies, showing good compatibility.

\begin{figure}[hb]
	%	\captionstyle{centerlast}
	\centering
	\includegraphics[width=.85\linewidth]{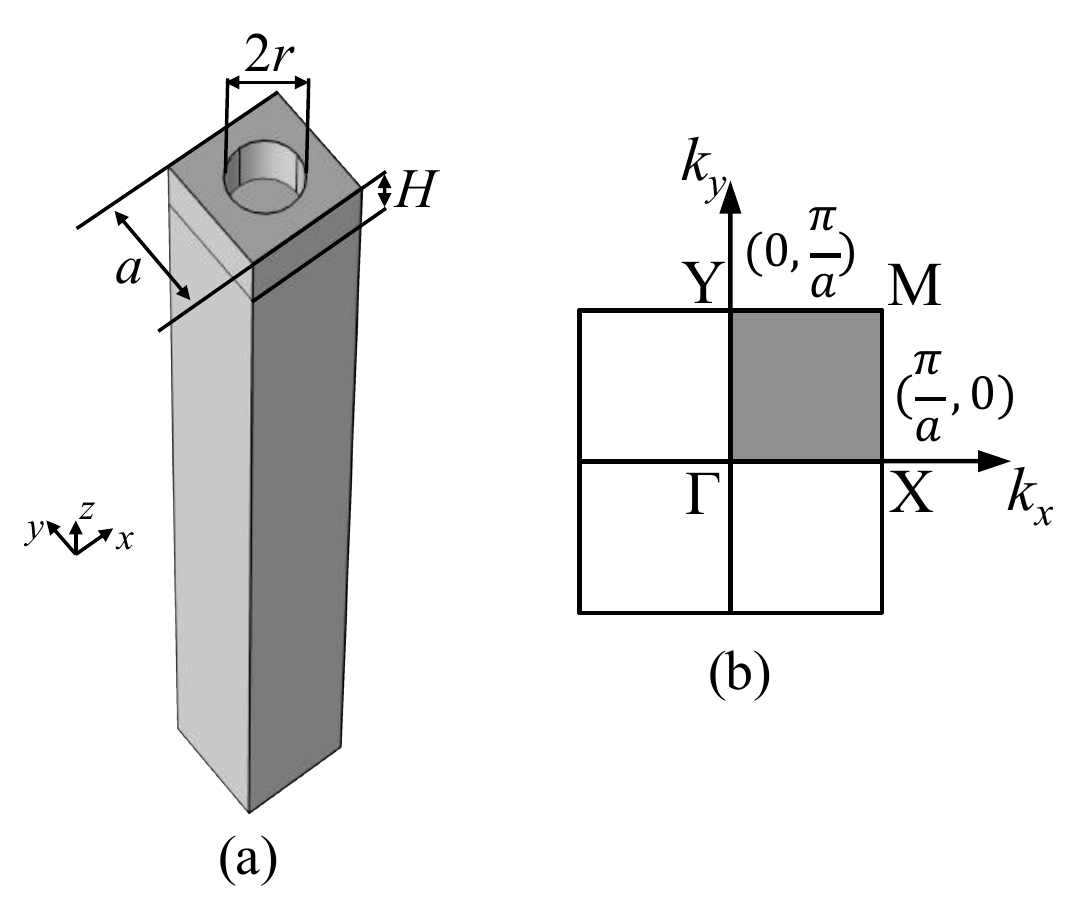}
	\caption{(a) Unit cell of the PnC with cylindrical holes arranged in square array in the silica film. The substrate is 90ST-cut quartz. $r=0.3a$, $H=0.6a$, $a=4\mu\text{m}$; (b) 1$^{st}$ BZ of the PnC. The gray square is the irreducible BZ; }
	\label{UC}
\end{figure}
\begin{figure*}[]
	\centering
	\includegraphics[width=0.9\linewidth]{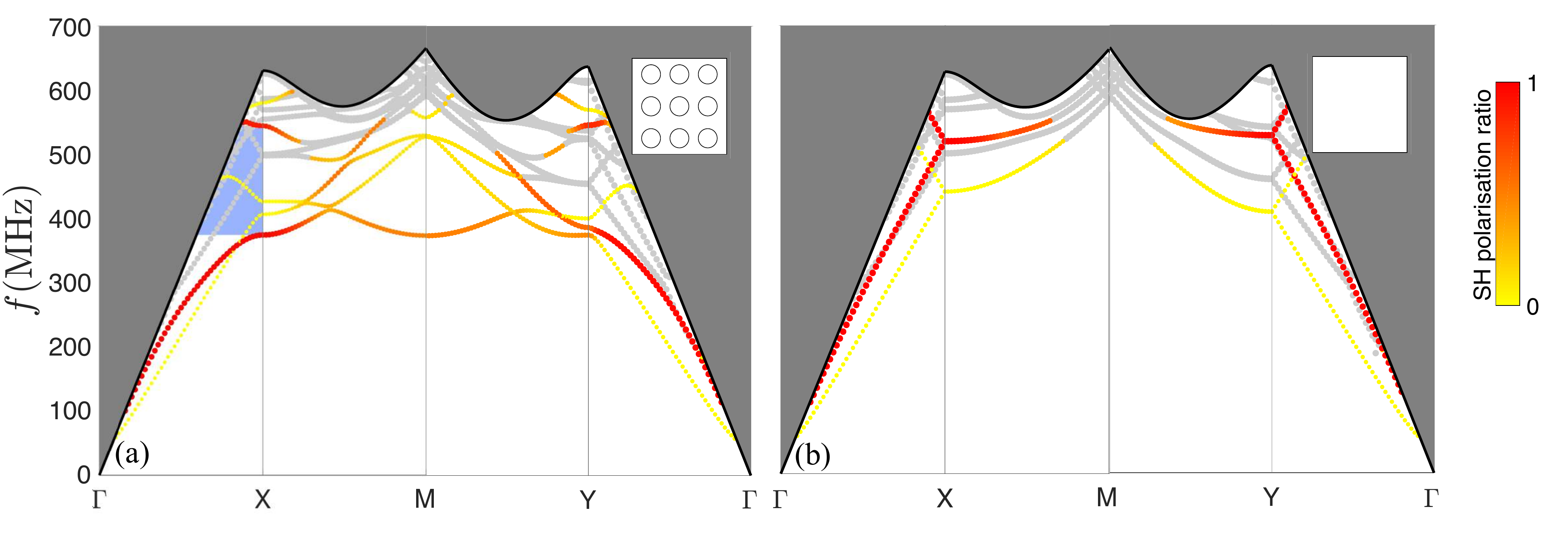}
	\caption{Complete band structures of (a) holey square arrayed PnC and (b) homogeneous silica on quartz (without PnC). Red-yellow colors denote the SH polarization ratio. Red indicates Love modes and yellow represents Rayleigh modes. Gray lines denote the modes propagating into the volume. Blue zone is the band gap in the $\Gamma$-X direction. $r=0.3a$, $H=0.6a$, $a=4\mu\text{m}$.}
	\label{bandstruct}
\end{figure*}

\section{Models and Simulation}
The guiding layer of Love waves is silica ($\rho=2200$kg/m$^{3}$,$E=70\textrm{GPa}, \nu=0.17$), with a height of $H=2.4\mu\textrm{m}$, covering a $40\mu\textrm{m}$-height 90ST-cut quartz substrate (Euler angles=(0\textdegree, 47.25\textdegree, 90\textdegree), LH 1978 IEEE), which has been rotated 90 degrees around the $z$-axis from the ST-cut quartz, for a fast SH waves (5000$\textrm{m/s}$) can be generated by the electric field.
The shear wave velocity in the silica film is 3438$\textrm{m/s}$, less than that in the 90ST-cut quartz substrate, indicating the existence of Love waves. The cylindrical holes in the silica film have a radius of $r=1.2 \mu\textrm{m}$. The square array period or the lattice constant is $a=4 \mu\textrm{m}$. The air hole is chosen because of its strong contrast in density and elastic constants with regard to the silica. A unit cell of the PnC constructed in COMSOL, showing in Fig~\ref{UC}(a), is employed to calculate the dispersion curves or the band structure. 
Floquet periodic boundary conditions are applied along the x and y directions to form a whole crystal, and the bottom of the substrate is assumed fixed. Love waves propagate along the $x$-axis (the $y$-axis of the ST-cut quartz), where Rayleigh waves can not be generated\cite{liu_design_2014} due to a zero electromechanical coupling factor to the substrate. 
%Note that the standard Euler angles, used for cutting the plate from a crystal, indicate the orientation of the plate relative to the crystal, while COMSOL defines the orientation of the crystal relative to the plate. Therefore the Euler angles of rotated system for this 90ST-cut quartz in COMSOL is (90\textdegree, 132.75\textdegree, 0\textdegree). %The substrate can also be a ST quartz (Euler angles=(0\textdegree, 47.25\textdegree, 0\textdegree) for cutting and (0\textdegree, 132.75\textdegree, 0\textdegree) in COMSOL) with the waves propagating along the $y$-axis. This means a rotation of 90\textdegree\ around the $z$-axis and the only thing to ensure is that the propagation stays in the direction where only the Love waves can be excited by the generated electric field. 
The first Brillouin zone (BZ) of the PnC is shown in Fig~\ref{UC}(b). Considering the anisotropy of the quartz substrate, the irreducible BZ is a square bounded by $\Gamma$-X-M-Y-$\Gamma$.
% However, the $\Gamma$-Y part is not of interest as it is supposed to reduce the coupling of the Rayleigh waves with the substrate. The band structure will hence be calculated on the triangular bounded by $\Gamma$-X-M-$\Gamma$.
The surface of the PnC coincides with the plane $z=0$. The wavelength normalized energy depth (NED) is calculated to select the surface modes for which is less than 1.  
\begin{equation}
\textrm{NED}=\frac{\iiint_{\mathcal{D}}\frac{1}{2}T_{ij}S_{ij}^{*}(-z)dxdydz}{\lambda\iiint_{\mathcal{D}}\frac{1}{2}T_{ij}S_{ij}^{*}dxdydz}
\end{equation}
$T_{ij}$ is the stress and $S_{ij}$ the strain. The asterisk (*) signifies the complex conjugate. $\mathcal{D}$ denotes the whole domain of the unit cell. $\lambda$ is the wavelength. Note that the integral in the denominator is the total acoustic potential energy in the unit cell and that the integral in the numerator is weighted by the depth of the point where the acoustic energy is not zero. That means if the average depth of the energy is less than the wavelength, the NED will be less than 1. The NED can well select the modes with speed less than the SH wave velocity of the substrate, where the wave vector $k$ is relatively large. As for a relatively small $k$, $\lambda$ is fixed to $2a$ that is resulting from $k=\frac{\pi}{a}$ and $k=\frac{2\pi}{\lambda}$. Moreover, the NED can filter out the plate modes appeared in our finite-depth substrate which is supposed to be semi-infinite for Love waves.

\begin{figure}[]
	\centering
	\includegraphics[width=1\linewidth]{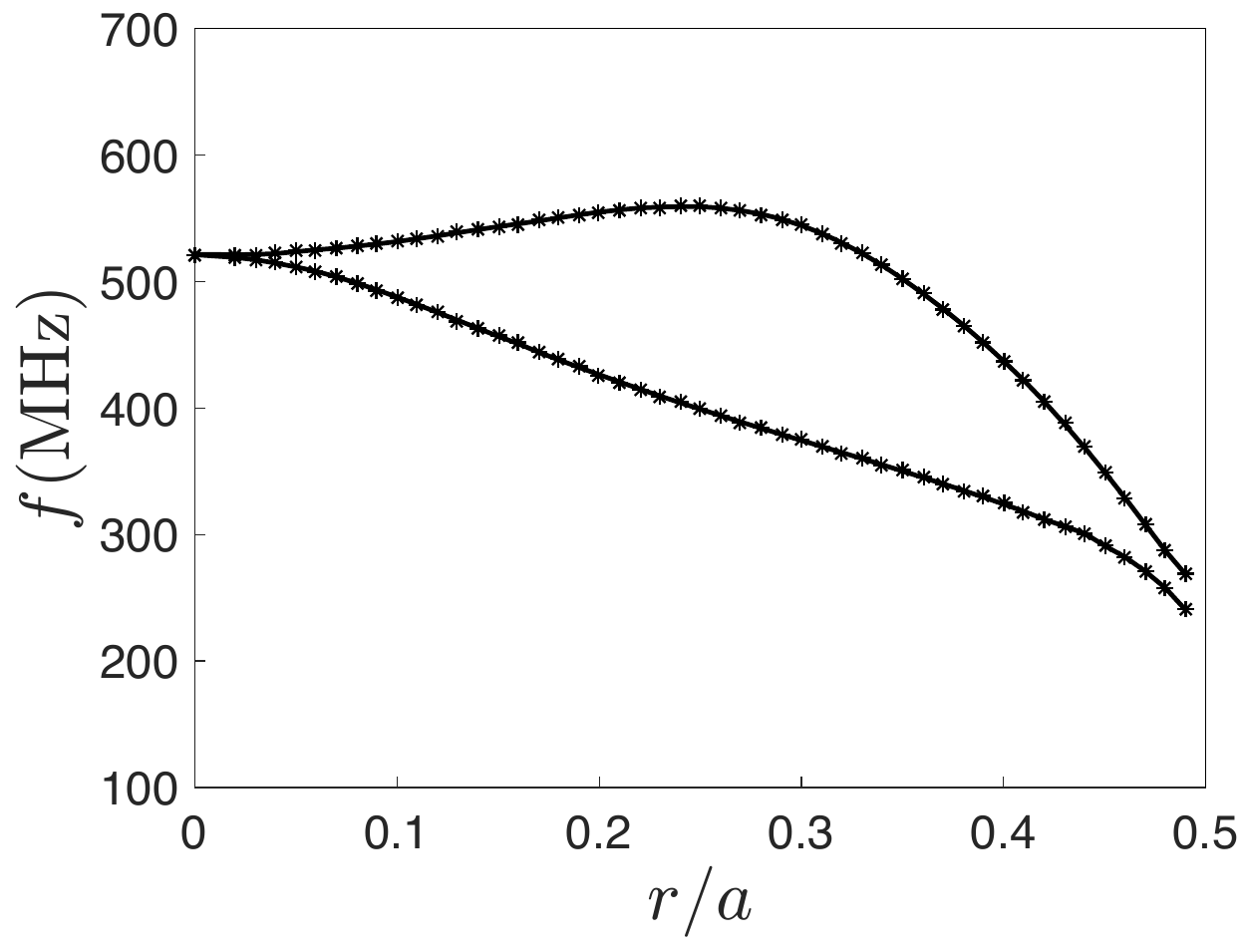}
	\caption{Love modes eigenfrequencies of the PnC as a function of the hole radius. Between the two black curves representing the two Love modes is the band-gap region of the PnC. $H=0.6a$, $a=4\mu\text{m}$.}
	\label{f-r}
\end{figure}

\begin{figure}[]
	\centering
	\includegraphics[width=1\linewidth]{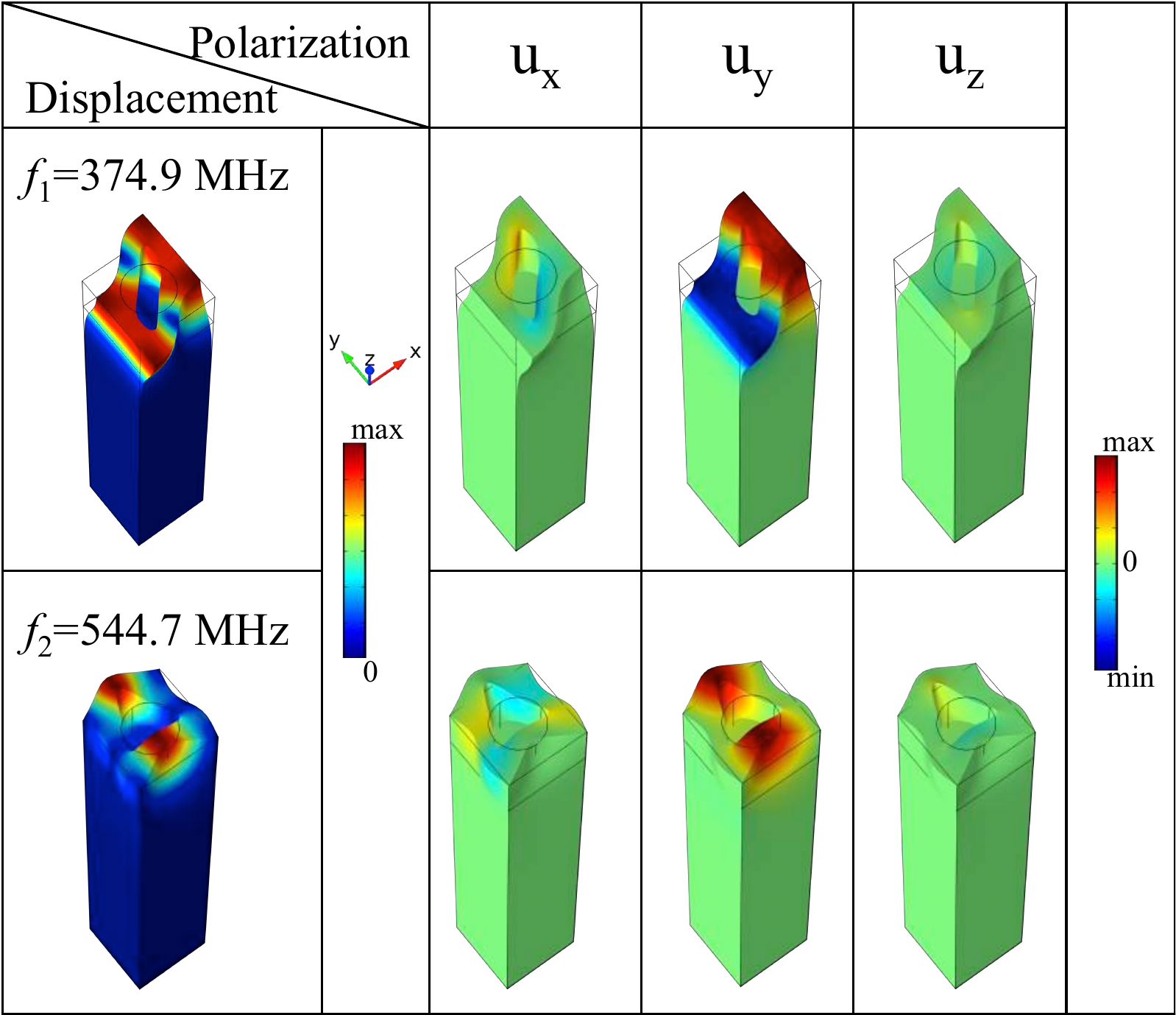}
	\caption{Displacement fields and polarizations of the two Love modes at 374.9 and 544.7 MHz. $r=0.3a$, $H=0.6a$, $a=4\mu\text{m}$.}
	\label{Polaris}
\end{figure}

Surface modes include SH type SAW and Rayleigh type SAW. The ratio of SH polarization is calculated to distinguish between Love waves and Rayleigh waves. 
\begin{equation}
\textrm{SH ratio}=\frac{\iiint_{\mathcal{D}}u_{SH}u_{SH}^{*}dxdydz}{\iiint_{\mathcal{D}}(u_{x}u_{x}^{*}+u_{y}u_{y}^{*}+u_{z}u_{z}^{*})dxdydz}
\end{equation}
$u_{x}, u_{y}$ and $u_{z}$ are respectively the displacements along the $x, y, z$ directions. $u_{SH}$ is the SH displacement component that can be expressed as $u_{x}\cos\theta-u_{y}\sin\theta$, which is perpendicular to the wave vector $\bm{k}$. $\theta$ is the angle between $\bm{k}$ and the $y$-axis with $\tan\theta=\frac{k_{x}}{k_{y}}$. The complete band structure calculated with COMSOL is shown in Fig~\ref{bandstruct}(a). The gray part is the radiation zone, where the waves diffuse to the volume (the bulk waves). The black line is the dispersion relation of the SH waves (here the fast shear waves) in the substrate, according to $v=\frac{2\pi f}{k}$. The curves in red and yellow denote the surface polarization modes. With the change of propagation direction, certain modes become gray as they start to diffuse into the volume. The modes colors are determined by their SH ratio. The red modes have a large SH ratio, indicating the Love modes. The yellower the modes, the closer they are to the Rayleigh type. Orange implies a coupling between Love modes and Rayleigh modes. In the $\Gamma$-X direction, the Love waves are not coupled to the Rayleigh waves, showing a large band gap ranging from 374.9 to 544.7 MHz between the two Love modes. Band structure without PnC is shown in Fig~\ref{bandstruct}(b), with no band gap for the Love modes. The relation between the band gap width and the normalized hole radius is shown by the two black curves in Fig~\ref{f-r}, where the band gap is between the two curves representing the two Love modes, consistent with the results reported by LIU et al.\cite{liu_evidence_2014}. The PnC gives rise to the appearance of the band gap, and the band width reaches a maximum (170.3 MHz) at $r/a$=0.29. Since the center of the band gap tends to decrease with the augmentation of the normalized radius, the relative band width ($\Delta f/f_{center}$) reaches its maximum (37.1\%) at $r/a$=0.31.
The displacement fields of the two Love modes at the point X of the BZ and their polarizations along the three axis are shown in Fig~\ref{Polaris}. It is found that the polarizations along the $x$-axis and the $z$-axis are negligible compared with the polarization on the $y$-axis, which means the polarizations of the two modes are horizontally perpendicular to the wave propagation direction ($x$-axis), proving that they are of SH type. Due to the exclusive generation of Love waves by the generated electric field, we only consider the Love modes in the rest of this paper. As the wave vector deviates from the $x$-axis, the band gap closes in the center of X-M. Therefore the propagation direction for calculating the transmission is along the $x$-axis. 

\begin{figure}[h!]
	\centering
	\includegraphics[width=1\linewidth]{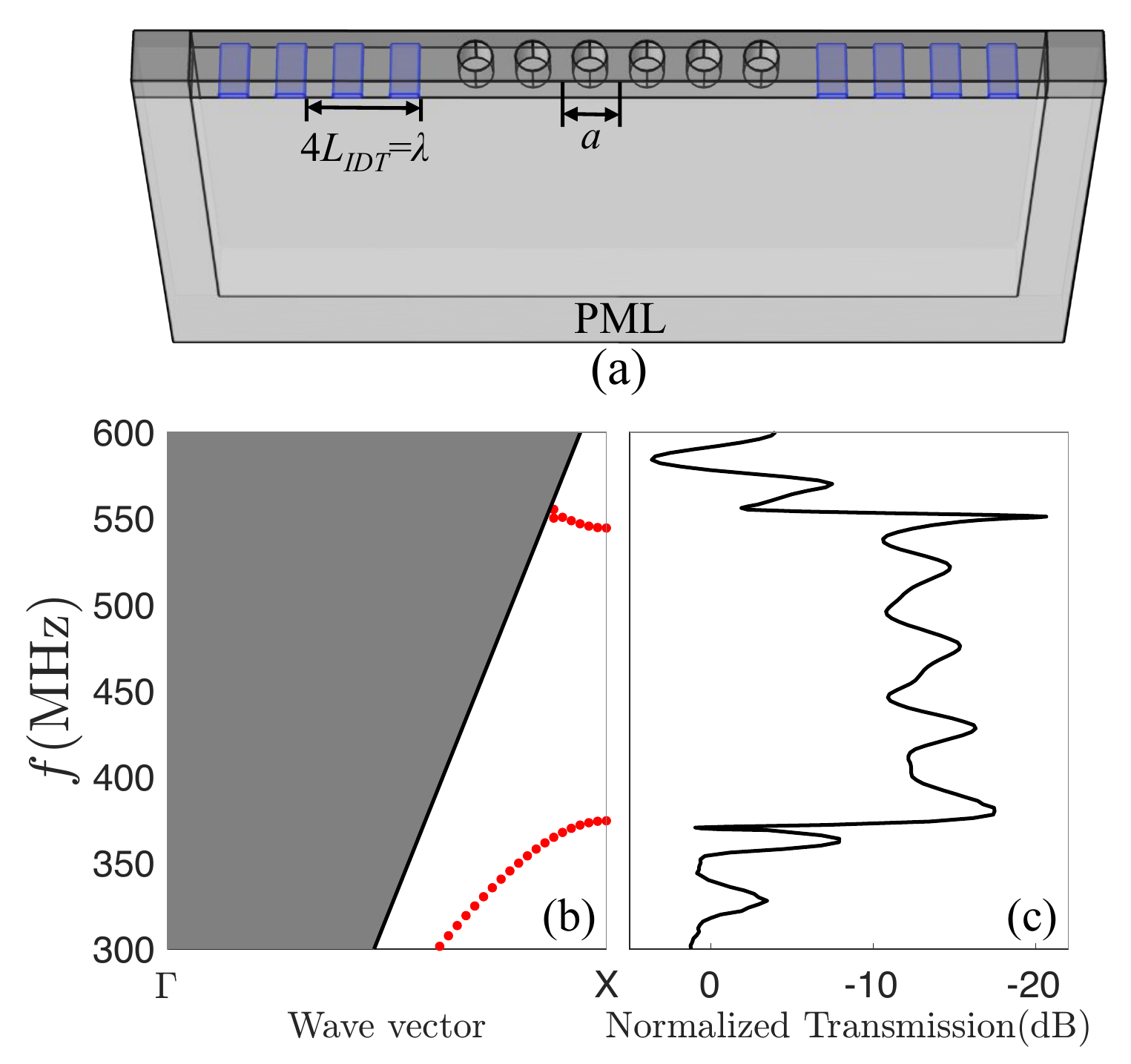}
	\caption{(a) Schematic of the SAW device model for calculating the transmission spectra of Love waves through the PnC. $r=0.3a$, $H=0.6a$, $a=4\mu\text{m}$; (b) Zoom of the band structure of Love modes around the band-gap zone in the $\Gamma$-X direction; (c) Normalized transmission spectra of Love waves propagating through the PnC around the band-gap zone. $N_{PnC}=10$, $N_{IDT}=20$, $h_{IDT}=200\textrm{nm}$, $V_{0}=1V$.}
	\label{TrUC}
\end{figure}
\begin{figure}[]
	\centering
	%	\begin{subfigure}[]{.35\linewidth}
	%		\centering
	\includegraphics[width=1\linewidth]{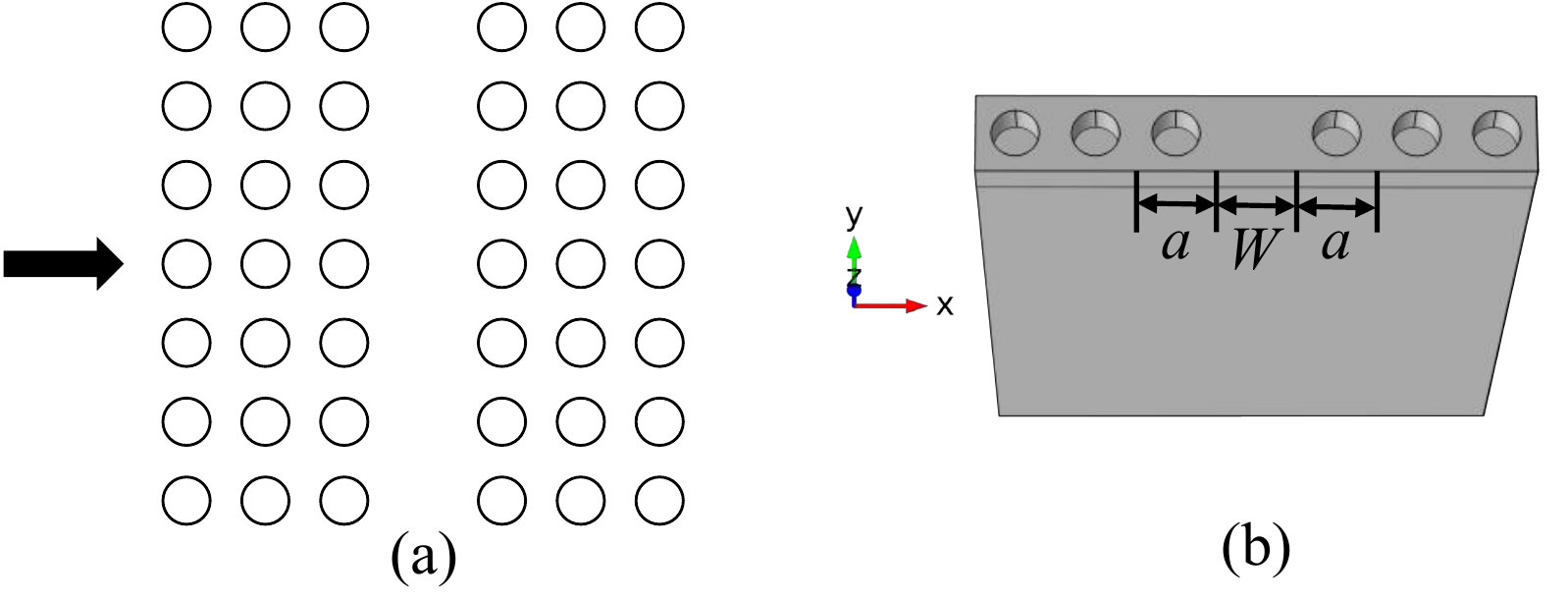}
	%\label{}
	%	\end{subfigure}   
	%	\begin{subfigure}[]{.55\linewidth}
	%		\centering
	%		\includegraphics[width=1\linewidth]{image/cavbandstruct}
	%		%\label{}
	%	\end{subfigure}
	\caption{(a) Schematic diagram of the PnC lattice containing a defect (cavity). The arrow denotes the direction of waves propagation; (b) Supercell of the defect-included PnC containing 6+$W/a$ unit cells, with $N_{PnC}=3$ on each side of the cavity. 
	}
	\label{BGCav}
\end{figure}

\begin{figure*}[]
	\centering
	\includegraphics[width=1\linewidth]{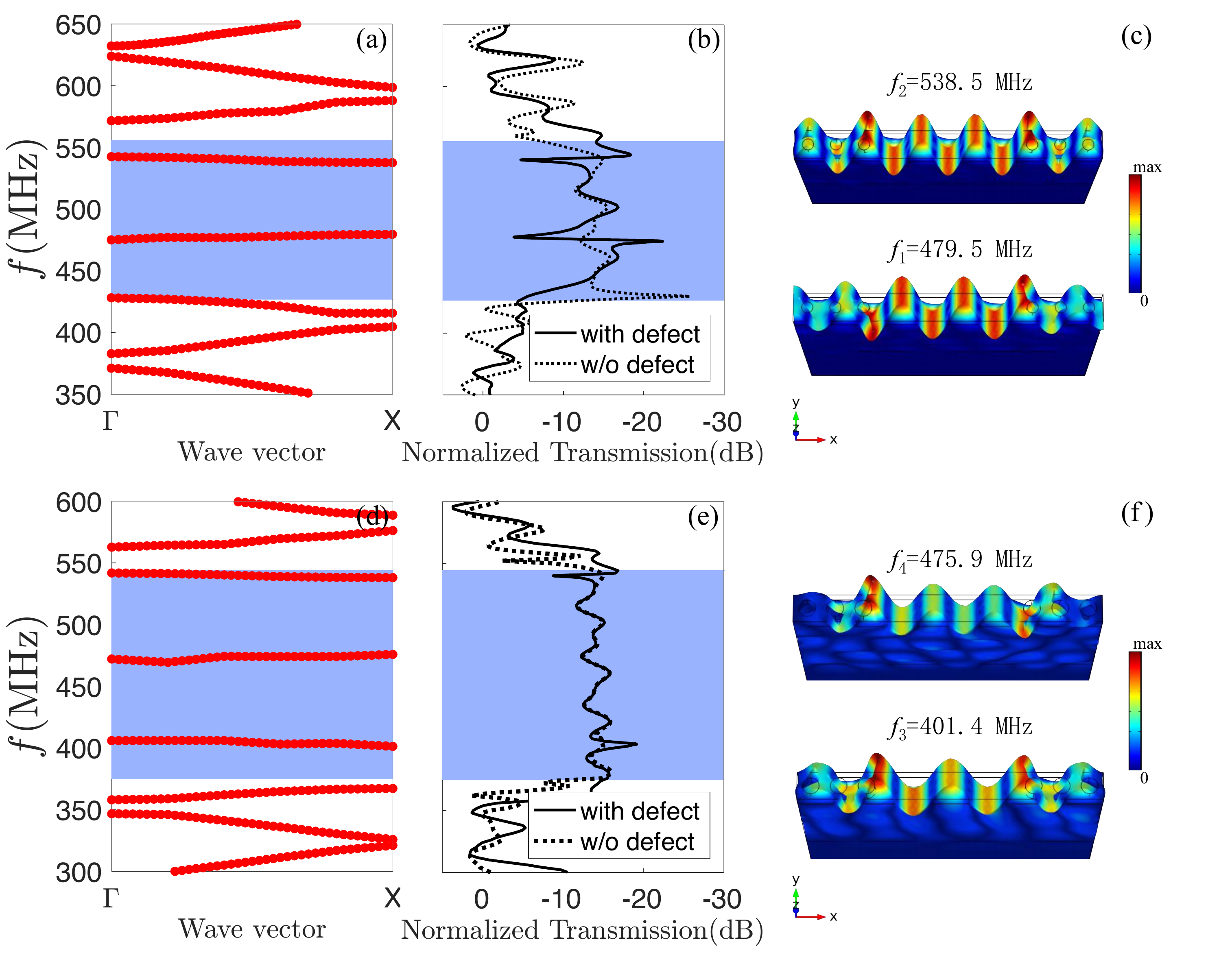}
	\caption{Band structures of Love modes on the defect-containing PnCs in the $\Gamma$-X direction around the band-gap region with (a)$r=0.2a$ and (d)$r=0.3a$, Blue zones are the predicted band gaps of the perfect PnCs; Normalized transmission spectra of Love waves with and without the defect (cavity) for (b)$r=0.2a$ and (e)$r=0.3a$, with $N_{PnC}=4$; Displacement fields of the supercells at the resonant frequencies of the cavity modes for (c)$r=0.2a$ and (f)$r=0.3a$. $W=5a$, $H=0.6a$, $a=4\mu\text{m}$}
	\label{bsTrW5}
\end{figure*}

\begin{figure*}[]
	\centering
	\includegraphics[width=1\linewidth]{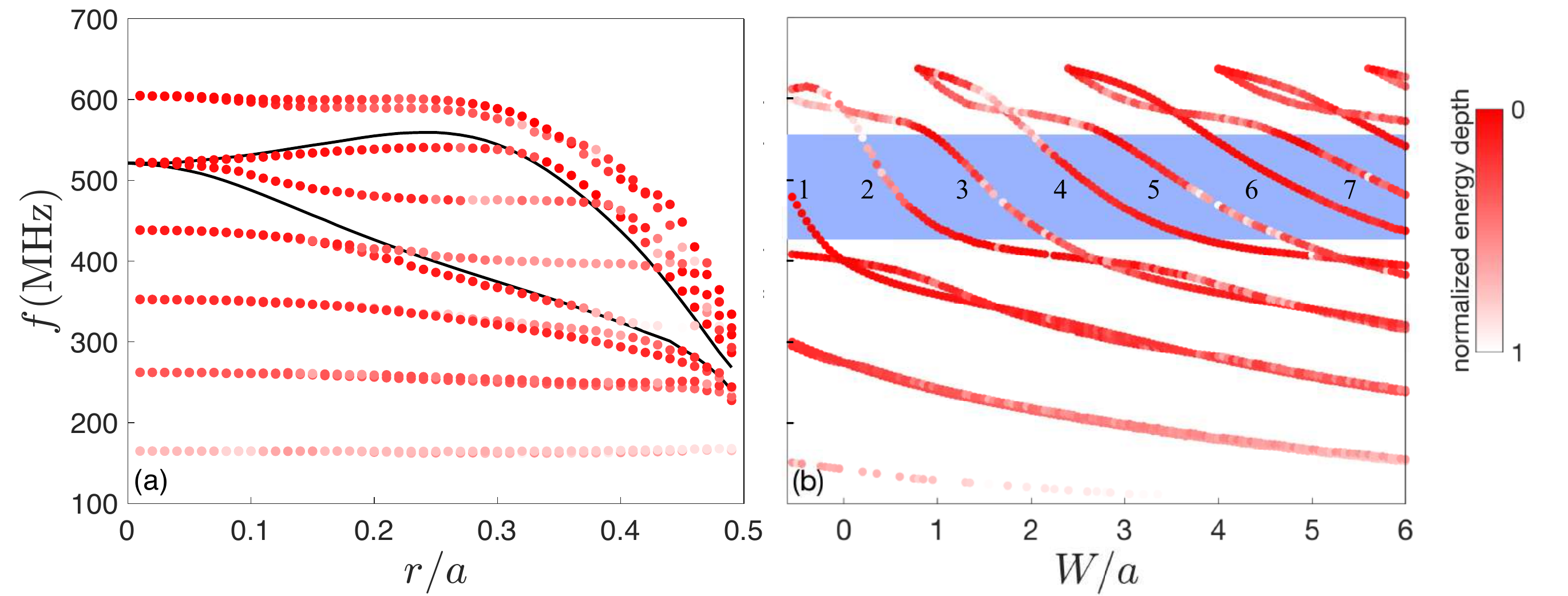}
	%	\begin{subfigure}[]{.49\linewidth}
	%		\centering
	%		\includegraphics[width=1\linewidth]{image/depth-r}
	%		%\label{}
	%	\end{subfigure}   
	%	\begin{subfigure}[]{.49\linewidth}
	%		\centering
	%		\includegraphics[width=1\linewidth]{image/depth-W}
	%		%\label{}
	%	\end{subfigure}
	\caption{(a) Love modes eigenfrequencies of the defect-containing PnC as a function of the hole radius. Inside the two black curves delimiting the band-gap zone of the perfect PnC are the cavity modes. Red-white colors denote the normalized energy depth (NED) of the Love modes. Red indicates a good confinement to the surface. $W=5a$, $H=0.6a$, $a=4\mu\textrm{m}$, $N_{PnC}=3$; (b)Love modes eigenfrequencies as a function of the cavity width. Blue zone is the predicted band gap of the perfect PnC. The numbers denote the order of the cavity modes. $r=0.2a$, $H=0.6a$, $a=4\mu\textrm{m}$, $N_{PnC}=3$.}
	\label{depth}
\end{figure*}

%\begin{figure}[hb]
%	\centering
%	\includegraphics[width=1\linewidth]{image/TrW3}
%	\caption{Transmission spectra for the PnC with defect with $W=3a$. $r=0.2a$, $H=0.6a$, $a=4\mu\text{m}$. Black dotted lines are the transmission spectrum of the perfect PnC. Blue zone is the predicted band gap of the perfect PnC. Red dotted lines are the resonant frequencies of predicted cavity modes in Fig~\ref{depthCav}(b).}
%	\label{TrW3}
%\end{figure}

\begin{figure}[b]
	\centering
	\includegraphics[width=1\linewidth]{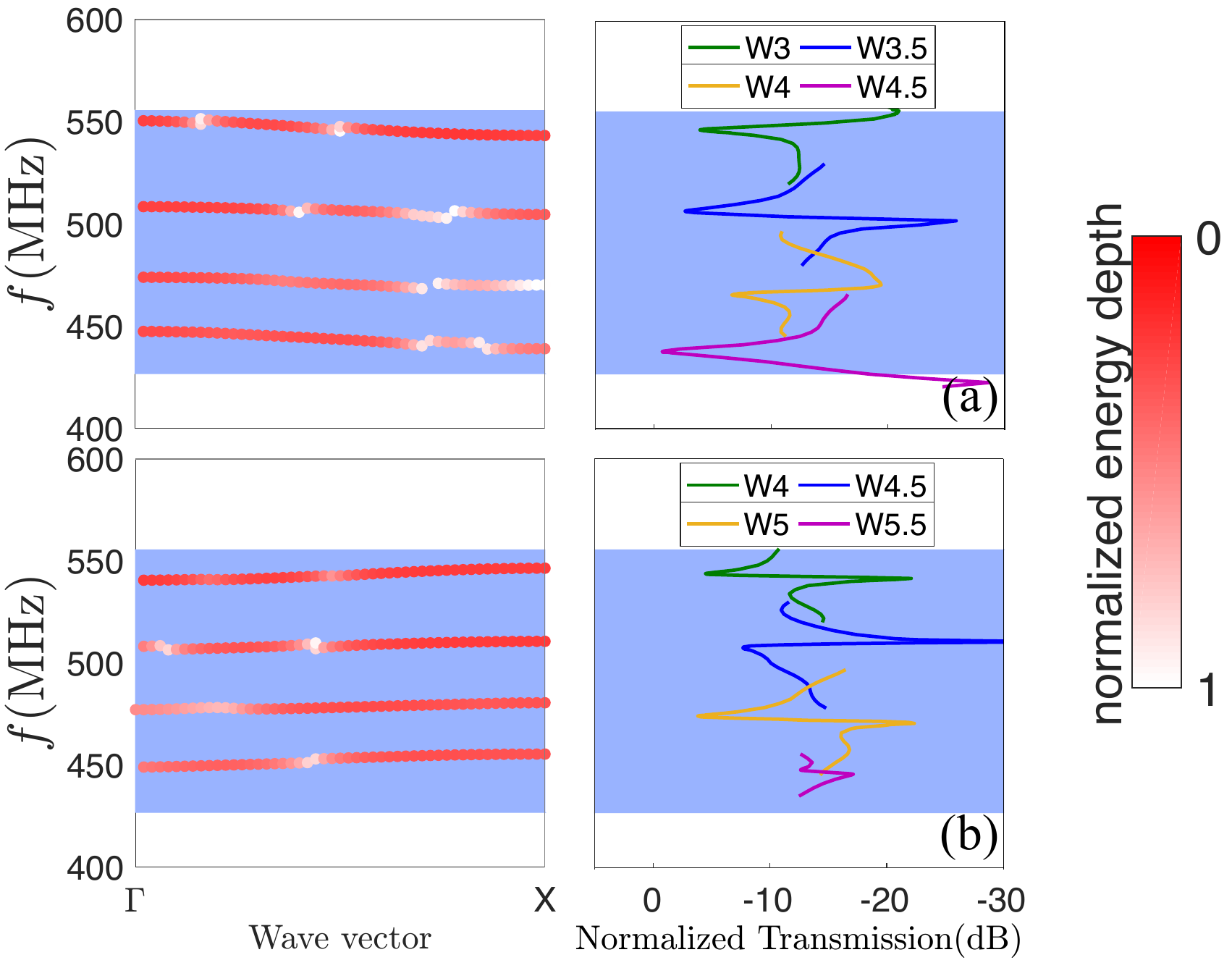} 
	%	\begin{subfigure}[]{.46\linewidth}
	%		\centering
	%		\includegraphics[width=1\linewidth]{image/TrM6}
	%		%\label{}
	%	\end{subfigure}
	\caption{Normalized transmission spectra for the defect-included PnC on the (a) 5$^{th}$ and (b) 6$^{th}$ cavity mode. Corresponding dispersion curve for each peak is shown on the left. Blue zones denote the predicted band gap of the perfect PnC. $r=0.2a$, $H=0.6a$, $a=4\mu\textrm{m}$, $N_{PnC}=4$. 
		%Red dotted lines are the resonant frequencies of cavity modes predicted by the supercell in Fig~\ref{depthCav}(b).
	}
	\label{TrW}
\end{figure}
\begin{figure}[b]
	\centering
	\includegraphics[width=0.9\linewidth]{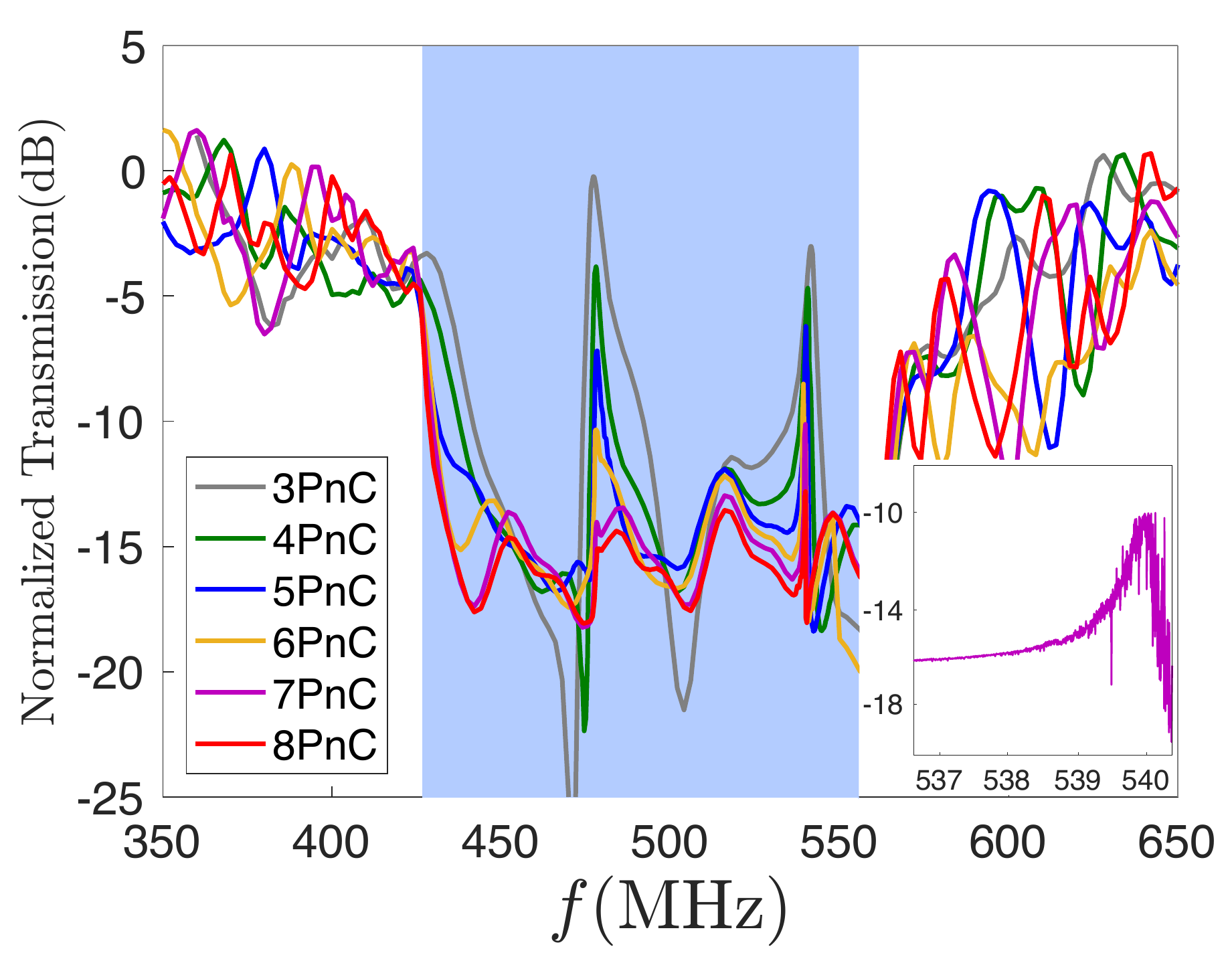}
	\caption{Transmission spectra of the PnC containing the defect, with the number of PnC holes on each side of the cavity varies from 3 to 8. Blue zone is the predicted band gap of the perfect PnC. Inset shows the zoomed peak at 540 MHz for $N_{PnC}=7$.
		$W=5a$, $r=0.2a$, $H=0.6a$, $a=4\mu\text{m}$.}
	\label{TrPC}
\end{figure}
\begin{figure}[]
	\centering
	\includegraphics[width=0.9\linewidth]{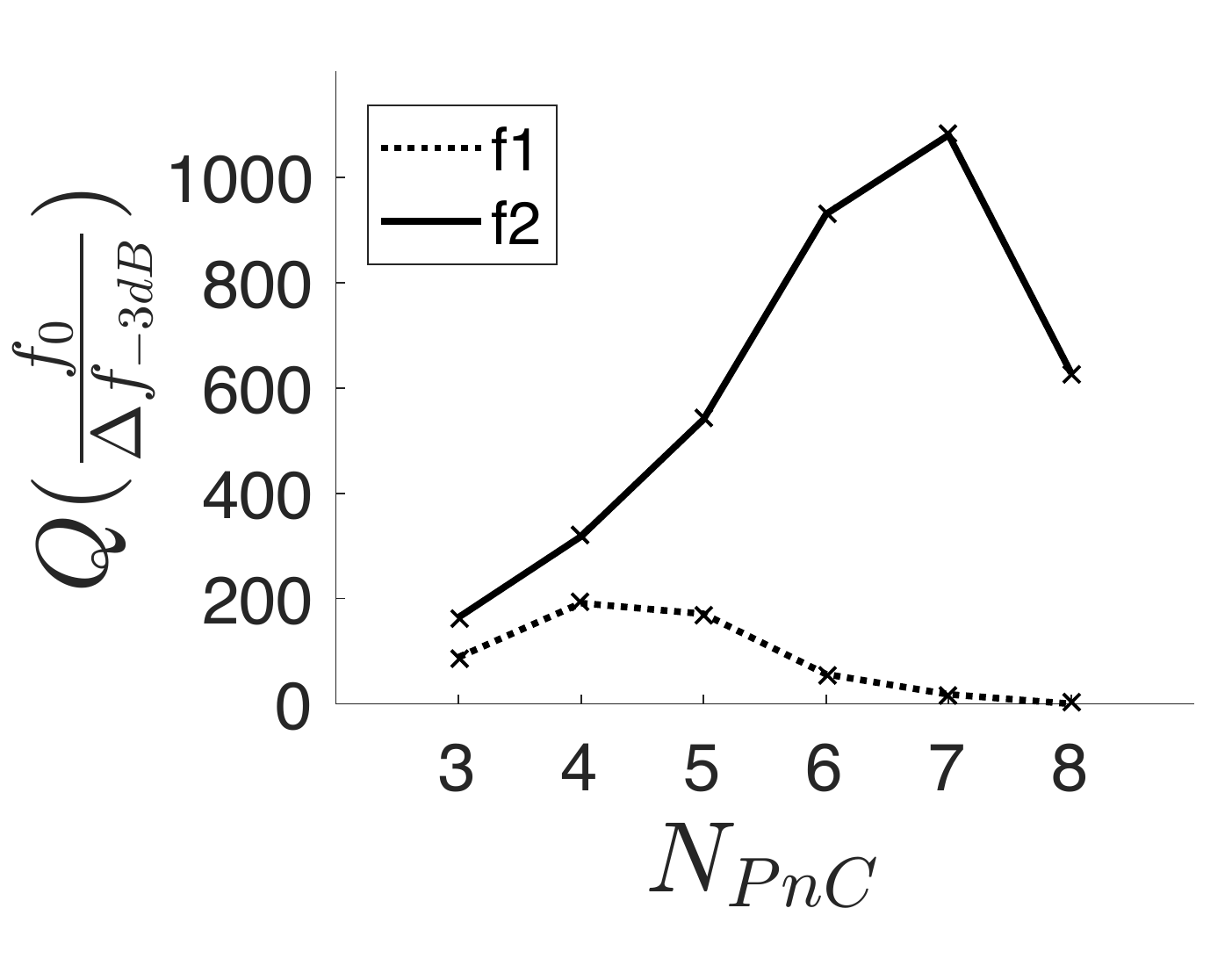}
	\caption{Quality factors of the defect modes as a function of the number of PnC holes on each side of the cavity.
		$W=5a$, $r=0.2a$, $H=0.6a$, $a=4\mu\text{m}$.}
	\label{Q}
\end{figure}

The calculation of transmission spectra is realized by simulating a SAW device consisting of two parts of aluminum inter-digital transducers (IDTs) and a PnC located between the IDTs. The height of IDTs is $h_{IDT}=200\textrm{nm}$. The IDTs are constructed on the quartz surface which is piezoelectric to generate the electric field. The graphic representation of the model is shown in Fig~\ref{TrUC}(a). Since the device has translational symmetry along the $y$-axis which is perpendicular to the direction of propagation, periodic boundary conditions are applied along the $y$-axis, reducing the simulation structure to only one period. The model is surrounded by perfectly matched layers (PMLs) for absorbing the undesired reflections from the boundary. The bottom and lateral sides are assumed fixed. One of the IDTs performing as a transmitter is given a harmonic voltage signal, with an amplitude of 1$V$, to excite acoustic waves in the quartz substrate. These waves are confined in the silica film and propagate through the PnC. They are received by the IDT on the other side. The output is measured by averaging the voltage difference between the odd and even fingers. The odd fingers of the input and output IDTs are assigned to the electrical ground. The even fingers of the output IDT are set to zero surface charge accumulation. Note that the width of the IDT fingers $L_{IDT}$ should be updated for each frequency in the spectrum according to the relation $L_{IDT}=\frac{\lambda}{4}=\frac{v}{4f}$. $v$ is the velocity of Love waves for $H=0.6a$, resulting from the basic dispersion relation of Love waves. That is, each frequency corresponds to a single wave velocity and wavelength. This frequency response is then normalized by that of the matrix (without PnC) to show the transmission loss contributed by the PnC only. Our previous work\cite{yankin_finite_2014} has proved the reliability of this model, giving coincident results between simulations and experiments. Fig~\ref{TrUC}(c) shows the normalized transmission spectra of the PnC, calculated with 10 PnC holes in center and 20 IDT fingers on each side. The attenuation appears clear and is consistent with the band structure prediction. %Fig~\ref{TrUC}(b) is the band structure in this direction around the band-gap zone.

The resonator is realized by removing $W$ lines of holes along the $y$ direction in the PnC lattice, forming a cavity perpendicular to the propagation direction. A supercell containing 1\texttimes ($6+W/a$) unit cells is constructed and shown in Fig~\ref{BGCav}(b), with periodic boundary conditions applied in the $x$ and $y$ directions. 
Here we set the cavity width to $W=5a$. 
The band structures of Love waves in the PnCs containing the defect are calculated along $\Gamma$-X and are shown in Fig~\ref{bsTrW5}(a) and (d), for $r/a$=0.2 and 0.3 respectively. 
Note that the $\Gamma$-X is 11 times smaller as we calculate for a supercell which is 11 times longer, and the band structure will be folded and repeated 11 times in the rest part. It is found that new flat modes, referred to as defect modes or cavity modes, appear inside the previously observed band gaps. These band structures are attributed to the coupling between the cavity and the perfect PnCs. Two cavity modes are predicted in Fig~\ref{bsTrW5}(a), respectively at 479.5 MHz and 538.5 MHz. In Fig~\ref{bsTrW5}(d), another two are at 401.4 and 475.9 MHz. 
The corresponding displacement fields of the cavity modes are shown in Fig~\ref{bsTrW5}(c) and (f). The displacements are concentrated in the center of the model (in the cavity) and attenuated at both ends. Inside the cavity, the displacements are uniform throughout the defect with maximums near the edges. It can be seen that the cavity modes for $r=0.2a$ are more confined to the surface than that for the radius of $0.3a$. In Fig~\ref{bsTrW5}(d), the flat mode on the upper limit of the band-gap region is not referred to as a cavity mode, since its displacement is no more concentrated in the cavity. 
Fig~\ref{bsTrW5}(b) and (e) show the normalized transmission spectra with and without the cavity in the PnCs, for the two different radius, calculated with 4 holes on each side of the cavity ($N_{PnC}$=4). For $r=0.2a$, two obvious peaks are found at 478 MHz and 540.6 MHz, consistent with the predicted resonant frequencies with small shifts due to the numeric mesh construction process of the FEM. These two flat cavity modes give rise to the highly confined transmission peaks. This means the cavity enables the propagation of waves that are otherwise forbidden in the perfect PnC. Each of the two transmission peaks possesses an antisymmetric line-shape, referred to as Fano resonance. The 1$^{st}$ transmission peak starts with an anti-resonance and ends with a resonance, while the 2$^{nd}$ transmission peak possesses the opposite behavior. 
%This is attributed to the opposite group velocity of the two cavity modes.
In Fig~\ref{bsTrW5}(e) for $r=0.3a$, only a small dip is found at 403.2 MHz, corresponding to the first defect mode in the band-gap region. This resonance is rather difficult to recognize from the band-gap bottom. Furthermore, no resonance has been found at the $2^{nd}$ predicted resonant frequency, which is referred to as a deaf mode. These phenomena might be resulting from the less confinement of the cavity modes for $r/a=0.3$. In other parts of the band-gap region, a superposition of the two curves is observed.

%The effects of the geometrical parameters on the cavity mode have been discussed as below to improve the performance of the resonator. 
Fig~\ref{depth}(a) shows the eigenfrequency-radius relation of a 5 holes removed supercell ($W$=5$a$), with 3 holes on each side of the cavity ($N_{PnC}$=3), calculated at the limit of the BZ (point X). Between the two black curves is the band-gap region of the perfect PnC. Two cavity modes are already in the band-gap region when r/a is near 0. They are separated with a certain distance as the radius increases. The two modes below penetrate into the band gap and become the cavity modes. It seems that the four cavity modes have a tendency to reach a similar distance from each other, referred to as mode spacing, the same way as the Love modes with r/a close to 0. Once this distance is reached, the lower external modes will cut in and their frequencies will be little affected by the normalized radius of the PnC. The modes outside are disturbed when approaching the band gap, and begin to surround this region. 
However, a larger radius of the holes and a lower order of the cavity modes result in a deeper mode energy (less confined to the surface), leading to a drop of energy transmission. The lowest cavity mode that appears around $r/a$=0.45 becomes even difficult to recognize. This explained the different transmission peaks for the $r=0.2a$ and $r=0.3a$ defect-containing PnCs. For this reason, we change the radius of the holes for our PnC to $r=0.2a$ in the rest of this paper, corresponding to a band gap extending from 426.8 to 555.5 MHz.

Fig~\ref{depth}(b) is the eigenfrequency-cavity width relation of a supercell with $r=0.2a$, calculated at point X. Inside the band gap denoted in blue are the cavity modes. As the cavity width increases, the frequencies of cavity modes decrease and modes with higher order appear (denoted by numbers). Apart the $1^{st}$ cavity mode, other modes are in pairs, and each pair is twisted outside the band-gap region and mutually merged. In our range of measurement (for $W$ from $-0.6a$ to $6a$), the confinement of Love modes in the band-gap region is better for a larger cavity width ($W>3a$) or a squeezed cavity width ($W<-0.3a$) . As the order of the cavity modes increases, the modes become less inclined, providing a larger cavity width range for each mode inside the band-gap region. For example, from 1$a$ to 2.3$a$ for the $3^{rd}$ cavity mode, and from 3$a$ to 4.7$a$ for the $5^{th}$ cavity mode. The wave period in the cavity increases by one-half for every higher mode order.

According to the cavity modes predictions in Fig~\ref{depth}(b), the transmission peaks of cavity modes can be displaced inside the band-gap region by changing the width of the cavity. The fifth and sixth cavity modes are shown as examples in Fig~\ref{TrW}. As the cavity width increases, the resonant frequency of each cavity mode decreases, with different occurring order of resonance and anti-resonance on the transmission peaks.
This proved the possibility of manipulation on the position of transmission peaks.
However, significant changes in shape are observed. This can be explained after carefully observing the dispersion curve of each peak.
It is found that the group velocity is negative for the 5$^{th}$ cavity mode and is positive for the 6$^{th}$ cavity mode. This is not influenced by the change in cavity width. However, the continuity and homogeneity of the dispersion curves are altered. A well confined (denoted in red) and smooth dispersion curve gives rise to a high and sharp transmission peak, see the cases of $W=4a$ and $W=5a$ for the 6$^{th}$ mode in Fig~\ref{TrW}(b). If the mode is leaky (denoted in white) somewhere in the dispersion curve and causes the curve to break (discontinuity or non-smoothness), the corresponding transmission peak might be affected, in terms of height and/or sharpness. The transmission peak of the 6$^{th}$ mode is perturbed on a $W=5.5a$ cavity, exhibiting a shift at the resonant frequency compared to the dispersion curve, which might be another explication of its shortened peak. A confined mode at point X is hence only a prerequisite for a high and sharp peak.
The dispersion curves for the 6$^{th}$ cavity mode are more compact than the curves for the 5$^{th}$ mode, as it possesses a larger range of cavity width within the band gap.
%The height of transmission peaks of a mode can be different inside the band gap, which is related to the band structure form. 

Fig~\ref{TrPC} shows the influences of the number of PnC holes (the crystal size) on the formation of peaks. As $N_{PnC}$ augments, the band-gap effect increases and it becomes harder for the waves to penetrate through the crystal, so the cavity peaks become shorter and sharper. The 1$^{st}$ transmission peak (at 478 MHz) drops more quickly than the 2$^{nd}$ peak, due to the enhanced band-gap effect in the center of the band-gap region, and eventually disappears after $N_{PnC}$ exceeds 7. On the other hand, insufficient crystal size ($N_{PnC}$ below 4) results in the blunt peaks, i.e., insufficient to filter out waves in the band-gap range other than the cavity frequencies.
The changes in quality factors (Qs) of the two peaks are shown in Fig~\ref{Q}. As $N_{PnC}$ augments, the Qs have a tendency to increase and then decrease. The Q of the 1$^{st}$ resonant mode decreases earlier due to its rapidly shortened peak. The Q of the 2$^{nd}$ peak reaches a maximum (1100) at $N_{PnC}=7$, where a highly confined cavity mode appears at 540 MHz, shown in the inset of Fig~\ref{TrPC}. It begins to decrease at $N_{PnC}=8$, owing to the shortened peak. 
For this reason, the crystal size should be properly chosen to keep a good quality factor as well as an isolation of the cavity modes. It is further confirmed that in our range of measurement ($3\le N_{PnC}\le 8$), the eigenfrequencies of cavity modes are independent of the crystal size, with only slight frequency shifts on the transmission peaks, which is mainly due to the mesh construction of the FEM.

%mode in center is smaller%

%Fig~\ref{TrW} shows the transmission spectra for different cavity width where the location of the cavity modes is different inside the band gap. The cavity transmission peaks are consistent with the resonant frequencies predicted by the supercell in Fig~\ref{depthCav}(b). Moreover, the peaks close to the band-gap edge are likely to be connected, as the first cavity mode in Fig~\ref{TrW}(a), with outer passing band. Although these cavity modes are all red in Fig~\ref{depthCav}(b), meaning that they are all highly confined at the surface, the transmission peak drops as the cavity mode comes to the center of the band gap, which is reasonable due to the band-gap effect of the PnC on the two sides. These dropped peaks could be increased by decreasing the number of PnC holes on sides of the cavity, as previously shown in Fig~\ref{TrPC}. However, the number of holes should be properly chosen as to keep a good quality factor and an isolation of the cavity modes.

\section{CONCLUTION}
In summary, we have presented the evidence of a partial band gap for Love waves propagating in a PnC consisting of holey square arrayed silica film on a 90ST-cut quartz. Cavity modes in the phononic band gap for Love waves are first demonstrated by removing lines of holes from the guiding film. The transmission peaks of cavity modes in the band gap of the perfect PnC is attributed to the appearance of new flat modes in the band structure. The transmission spectra are proved to be consistent with the band structure predictions. The resonant frequencies of cavity modes are related to the cavity width. A well confined and flat cavity mode, as well as a properly chosen PnC size, is essential for obtaining sharp transmission peaks. This study could be used for potential applications of Love wave-based PnC devices.

\appendix
%\nocite{*}
%\bibliography{PnC}% Produces the bibliography via BibTeX.
%\printbibliography

\end{document}